\documentclass[english]{IEEEtran}
\usepackage[T1]{fontenc}
\usepackage[latin9]{inputenc}
\usepackage{float}
\usepackage{amsmath}
\usepackage{amsthm}
\usepackage{amssymb}
\usepackage{stackrel}
\usepackage{graphicx}

\makeatletter

\providecommand{\tabularnewline}{\\}

\theoremstyle{plain}
\newtheorem{thm}{\protect\theoremname}
\theoremstyle{definition}
\newtheorem{defn}[thm]{\protect\definitionname}
\theoremstyle{definition}
\newtheorem{example}[thm]{\protect\examplename}
\theoremstyle{plain}
\newtheorem{lem}[thm]{\protect\lemmaname}

\makeatother

\usepackage{babel}
\providecommand{\definitionname}{Definition}
\providecommand{\examplename}{Example}
\providecommand{\lemmaname}{Lemma}
\providecommand{\theoremname}{Theorem}

\begin{document}
\title{Multiple-Output Channel Simulation and Lossy Compression of Probability
Distributions}
\author{Chak Fung Choi and Cheuk Ting Li\\
The Chinese University of Hong Kong\\
Email: cfchoi@link.cuhk.edu.hk, ctli@ie.cuhk.edu.hk}

\maketitle

\begin{abstract}
We consider a variant of the channel simulation problem with a single
input and multiple outputs, where Alice observes a probability distribution
$P$ from a set of prescribed probability distributions $\mathbb{\mathcal{P}}$,
and sends a prefix-free codeword $W$ to Bob to allow him to generate
$n$ i.i.d. random variables $X_{1},X_{2,}...,X_{n}$ which follow
the distribution $P$. This can also be regarded as a lossy compression
setting for probability distributions. This paper describes encoding
schemes for three cases of $P$: $P$ is a distribution over positive
integers, $P$ is a continuous distribution over $[0,1]$ with a non-increasing
pdf, and $P$ is a continuous distribution over $[0,\infty)$ with
a non-increasing pdf. We show that the growth rate of the expected
codeword length is sub-linear in $n$ when a power law bound or exponential
tail bound is satisfied. An application of multiple-outputs channel
simulation is the compression of probability distributions. 
\end{abstract}

\section{Introduction}

\allowdisplaybreaks

The asymptotic channel simulation problem \cite{bennett2002entanglement,winter2002compression}
is described as follows. Let $\mathbb{\mathcal{P}}=\{P_{\theta}:\theta\in\mathcal{A}\}$
be a set of probability distributions indexed by $\theta$. The encoder
observes $\theta_{1},\ldots,\theta_{n}$ and sends a message $M$
to the decoder. The decoder then outputs $X_{1},\ldots,X_{n}$. The
encoder and decoder may also share common randomness. The goal is
to have the conditional distribution of $X_{1},\ldots,X_{n}$ given
$\theta_{1},\ldots,\theta_{n}$ to be approximately $P_{\theta_{1}}\times\cdots\times P_{\theta_{n}}$,
while minimizing the rate of the message $M$ as $n\to\infty$. It
was shown by Bennett et. al. \cite{bennett2002entanglement} that
for the case with unlimited common randomness, the optimal rate is
given by $C=\max_{p(\theta)}I(\theta;X)$ (where $X$ follows the
conditional distribution $P_{\theta}$ given $\theta$), i.e., the
capacity of the channel $\theta\to X$. For the case where $\theta$
is known to follow the distribution $p(\theta)$, Winter \cite{winter2002compression}
showed that a rate of $I(\theta;X)$ for the message, and a rate of
$H(X|\theta)$ for the common randomness suffices. Cuff \cite{cuff2013distributed}
characterized the optimal trade-off between the communication rate
and the common randomness rate.

The channel simulation problem is also studied in a one-shot setting
($n=1$), where the encoder observes $\theta$ and sends a prefix-free
codeword $W$ to the decoder, which then outputs $X$. The goal is
to have $X$ follows the conditional distribution $P_{\theta}$ given
$\theta$ exactly, while minimizing the expected length $\mathbf{E}(L(W))$
of $W$. Harsha et al. \cite{harsha2007communication} studied the
case with unlimited common randomness, and showed that $\mathbf{E}(L(W))\le C+(1+\epsilon)\log(C+1)+O(1)$
bits of codeword and $O(\log|\mathcal{X}|+\log|\mathcal{Y}|)$ bits
of common randomness suffices for one-shot setting (where $C$ is
the capacity of the channel $\theta\to X$, and $|\mathcal{X}|$ denotes
the cardinality of $X$). Braverman and Garg \cite{braverman2014public}
improved the result by eliminating the multiplicative factor $(1+\epsilon)$.
Li and El Gamal \cite{li2018strong} strengthened the result by showing
that $C+\log(C+1)+5$ bits of codeword and $\log(|\mathcal{X}|(|\mathcal{Y}|-1)+2)$
bits of common randomness suffice. The case without common randomness
is studied in \cite{kumar2014exact,li2017distributed}.

A universal setting where $\mathbb{\mathcal{P}}$ is the class of
continuous distributions over $\mathbb{R}$ was studied by Li and
El Gamal \cite{li2018universal}. In this case, it is more natural
to omit the index $\theta$ and assume the encoder observes a distribution
$P\in\mathbb{\mathcal{P}}$, and the expected length would depend
on $P$.

\begin{figure}[H]
\begin{centering}
\includegraphics[scale=0.44]{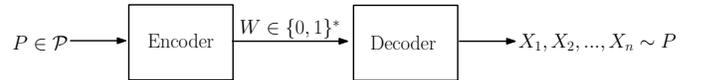}
\par\end{centering}
\caption{\label{fig:setting}Multiple-output channel simulation setting.}
\end{figure}

This paper studies an extension of the one-shot universal channel
simulation setting, called the multiple-output channel simulation
setting, described as follows. The encoder observes $P\in\mathbb{\mathcal{P}}$
and sends a codeword $W\in\{0,1\}^{*}$ from an agreed-upon prefix-free
code to the decoder. The decoder then outputs $X_{1},...,X_{n}$.
There is no common randomness shared between the encoder and the decoder.
The goal is to have $X_{1},...,X_{n}$ i.i.d. following $P$ exactly,
while minimizing the expected length $\mathbf{E}(L(W))$. This setting
is depicted in Figure \ref{fig:setting}. A straightforward approach
is to apply the scheme in \cite{li2018universal} $n$ times, resulting
in an expected length that grows linearly in $n$. In this paper,
we are interested in schemes where the expected length that grows
sublinearly in $n$. 

Another approach is to have the encoder generate $X_{1},...,X_{n}$
i.i.d. following $P$ and encode them into $W$, so the decoder can
decode $X_{1},...,X_{n}$. Since the decoder can perform a random
shuffle on its output, the ordering of $X_{1},...,X_{n}$ does not
matter, that is, the encoder only need to encode the multiset $\{X_{1},...,X_{n}\}$.
The problem of encoding multisets was studied by Varshney and Goyal
\cite{varshney2006ordered,varshney2006toward,varshney2007benefiting},
who showed that for the case where the alphabet $\mathcal{X}$ is
finite, $|\mathcal{X}|\log(n+1)$ bits suffice to encode the multiset
$\{X_{1},...,X_{n}\}$. Nevertheless, this approach is inapplicable
for the case where the space $\mathcal{X}$ is continuous, since it
is impossible to encode a real number into a finite number of bits.

In this paper, we study three cases of the class of distributions
$\mathcal{P}$, where an expected length that grows sublinearly in
$n$ is possible. In Section \ref{sec:p_int}, we present a scheme
for the case where $\mathcal{P}$ is the class of distributions over
positive integers. This scheme is also applicable to the problem of
encoding multisets \cite{varshney2006ordered,varshney2006toward,varshney2007benefiting}.
In Section \ref{sec:p_01}, we present a scheme for the case where
$\mathcal{P}$ is the class of continuous distribution over $[0,1]$
with a non-increasing pdf. Our scheme is based on the dyadic decomposition
construction in \cite{li2017distributed,li2018universal}. In Section
\ref{sec:p_inf}, we present a scheme for the case where $\mathcal{P}$
is the class of continuous distribution over $[0,\infty)$ with a
non-increasing pdf, which combines the two aforementioned schemes.

For an application of multiple-output channel simulation, consider
the setting of lossy compression of a probability distribution. The
encoder encodes $P\in\mathcal{P}$ into $W\in\{0,1\}^{*}$. The decoder
decodes $W$ into $\hat{P}$. For the case where $\mathcal{P}$ is
the class of continuous distributions over real numbers, one method
is to approximate the pdf of $P$ by a piecewise linear function $\hat{P}$
with vertices that have rational coordinates, and compress those coordinates
into $W$. There are two shortcomings of this method. First, the main
use of a probability distribution is to simulate random variables
from it, but it is impossible to obtain samples following $P$ exactly
using $W$ or $\hat{P}$ (we can only sample from $\hat{P}$ which
is inexact). Second, this method generally produces $\hat{P}$ that
is a biased estimate of $P$, i.e., $\mathbf{E}[\hat{P}(A)]\neq P(A)$
for some $A\subseteq\mathbb{R}$. More sophisticated kernel interpolation
techniques may be used to approximate $P$, but the same problems
persist.

Using multiple-output channel simulation, we can allow the decoder
to obtain i.i.d. samples $X_{1},...,X_{n}$ following $P$, and produce
the estimate as the empirical distribution $\hat{P}(A)=n^{-1}\sum_{i=1}^{n}\mathbf{1}\{X_{i}\in A\}$.
This overcomes the two aforementioned problems. First, the decoder
can obtain exact i.i.d. samples of $P$ as long as the number of samples
is not greater than $n$. Second, $\hat{P}$ is an unbiased estimate
of $P$, i.e., $\mathbf{E}[\hat{P}(A)]=P(A)$ for $A\subseteq\mathbb{R}$.
Our scheme allows the decoder to obtain i.i.d. samples and perform
statistical tests on $P$, without the need of transmitting all information
about $P$.

Throughout this paper, we assume that $\log$ is base 2. Log in base
$e$ is written as $\ln$. We use the notation: $\left[a:b\right]=\left[a,b\right]\cap\mathbb{Z}$.
For two bit sequences $A,B\in\{0,1\}^{*}$, denote their concatenation
as $A\Vert B$, and the length of $A$ as $L(A)$.

\section{$P$ is a distribution over positive integers\label{sec:p_int}}

This section develops a coding scheme for the case where $P$ is a
distribution over positive integers, called \emph{difference run-length
encoding scheme}. We then show that when $P$ satisfies the bound
$P(X>x)\leq cx^{-\lambda}$ or $P(X>x)\leq ce^{-\lambda x}$, where
$c,\lambda>1$, the expected codeword length for encoding is $o(n)$.

We first review the Elias gamma code \cite{elias1975universal}.
\begin{defn}
[Elias gamma code \cite{elias1975universal}] Let $Z$ be a positive
integer. The codeword $g(Z)$ is defined as
\begin{align*}
g(Z) & =0^{N}\Vert1\Vert a_{N-1}a_{N-2}...a_{0},
\end{align*}
where $a_{N}a_{N-1}...a_{0}$ is the binary representation of $Z$.
The length of the codeword $g(Z)$ is $L(g(Z))=\left\lfloor 2\log Z+1\right\rfloor .$
\end{defn}
We then define the difference run-length coding scheme.
\begin{defn}
[Difference run-length encoding scheme]\label{def:scheme_p_int}
The encoder and decoder are described as follows:
\begin{enumerate}
\item The operations of the encoder are:
\begin{enumerate}
\item Generate i.i.d. $\widetilde{X}_{1},\widetilde{X}_{2},...,\widetilde{X}_{n}\sim P$.
\item Sort $\widetilde{X}_{1},\widetilde{X}_{2},...,\widetilde{X}_{n}$
in ascending order such that $\widetilde{X}_{(1)}\leq\widetilde{X}_{(2)}\leq\cdots\leq\widetilde{X}_{(n)}$.
\item Let $D_{1}=\widetilde{X}_{(1)}.$ For $i\in[2:n]$, let $D_{i}=\widetilde{X}_{(i)}-\widetilde{X}_{(i-1)}$.
\item Let $w_{1}=g(D_{1})$. For $i\in[2:n]$, if $D_{i}>0$, then let $w_{i}=1||g(D_{i})$,
where the first bit ``1'' indicates that $D_{i}>0$. If $D_{i}=0$
and $D_{i-1}\neq0$, the encoder finds the smallest positive integer
$j_{i}$ such that $D_{i+j_{i}}\neq0$ (i.e., $j_{i}$ is the number
of consecutive zeros starting at index $i$; assume $D_{n+1}=1$),
then let $w_{i}=0\Vert g(j_{i}),$ where the first bit ``0'' indicates
that $D_{i}=0$ and $D_{i-1}\neq0$. If $D_{i}=0$ and $D_{i-1}=0,$
then let $w_{i}=\emptyset$ (the empty sequence). Then, the encoder
sends a codeword $W$, which is a series of concatenated $w_{i}$,
i.e., $W=w_{1}\Vert w_{2}\Vert...\Vert w_{n}$, to the decoder. 
\end{enumerate}
\item The operations of the decoder are:
\begin{enumerate}
\item Upon receiving $w$, the decoder uses the Elias gamma code to decode
$\widetilde{X}_{(1)}=D_{1}$, and discard the decoded bits. Initialize
$i=2$.
\item If the next undecoded bit is $1$, the decoder discard that bit, decode
$D_{i}$ (and discard the decoded bits), compute $\widetilde{X}_{(i)}=\widetilde{X}_{(i-1)}+D_{i}$,
and increment $i$. If the next undecoded bit is $0$, the decoder
discard that bit, decode $j_{i}$ (and discard the decoded bits),
compute $\widetilde{X}_{(i)},\widetilde{X}_{(i+1)},...,\widetilde{X}_{(i+j_{i}-1)}=\widetilde{X}_{(i-1)}$,
and increment $i$ by $j_{i}$. Repeat this step until there is no
more undecoded bit.
\item Lastly, it shuffles $\widetilde{X}_{(1)},\widetilde{X}_{(2)},...,\widetilde{X}_{(n)}$
randomly and outputs them as $X_{1},X_{2},...,X_{n}$.
\end{enumerate}
\end{enumerate}
\end{defn}
\medskip{}

\begin{example}
\label{exa:geom}Suppose $P$ is the geometric distribution $\mathrm{Geom}(0.7)$.
Alice generates 10,000 instances of $\widetilde{X}_{i}\sim\mathrm{Geom}(0.7)$,
with frequencies summarized in Table \ref{tab:example}. The codeword
$w$ will be in the form $g(1)\Vert0\Vert g(7039)\Vert1\Vert g(1)\Vert\cdots\Vert0\Vert g(2)\Vert1\Vert g(1)$.
Through direct computation, the codeword length is $139$, which is
significantly less than $n=10,000$.
\end{example}
\begin{table}
\begin{centering}
\begin{tabular}{|c|c|c|c|c|c|c|c|c|c|}
\hline 
$X_{i}$ & 1 & 2 & 3 & 4 & 5 & 6 & 7 & 8 & 9\tabularnewline
\hline 
Freq. & 7040 & 2056 & 641 & 184 & 53 & 13 & 9 & 3 & 1\tabularnewline
\hline 
\end{tabular}
\par\end{centering}
\centering{}\medskip{}
\caption{\label{tab:example}Table of frequencies of $\widetilde{X}_{i}$ and
$X_{i}$ for Example \ref{exa:geom}. }
\end{table}

Our method uses the difference of two consecutive integers to reduce
the encoded integer's magnitude, which reduce the length of the Elias
delta code. Also, we observe that the difference sequence contains
consecutive zeros. Therefore, we use the technique from run-length
encoding to reduce the length of codeword. Consequently, our coding
scheme can significantly reduce the codeword length for data that
concentrates on specific positive integers.

We present the following theorem, which shows that the codeword length
grows sub-linearly in $n$ when $P$ follows a power law bound.
\begin{thm}
\label{thm:p_int}Let $P$ be a distribution over positive integers.
If $P$ satisfies the bound $P(X>x)\leq cx^{-\lambda}$ for all integer
$x\ge0$, where $c>1$ and $\lambda>1$, then the expected codeword
length of difference run-length coding scheme for $P$ is upper bounded
as
\begin{align*}
\mathbf{E}(L(W)) & \leq\dfrac{50c\lambda n^{\frac{1}{\lambda}}\log(\sqrt{n}+1)}{\lambda-1}.
\end{align*}
\end{thm}
\begin{IEEEproof}
We will separate the set of indices into two parts and calculate an
upper bound of expected codeword length for encoding these two parts
separately. Consider the sets $U=\{i:D_{i}>0\}$, $V=\{i:D_{i}=0\mbox{ and }D_{i-1}\neq0\}$.
Note that $w_{i}\neq\emptyset$ only if $i\in U\cup V$.

Consider $i\in U$. Note that $L(w_{i})=\left\lfloor 2\log D_{i}+1\right\rfloor +1\leq2\log D_{i}+2\leq2\log(2D_{i}+1)$.
Define $\ell(x)=2\log(2x+1)$, which is a concave function. We have

\begin{align*}
 & \mathbf{E}\left[\sum_{i\in U}L(w_{i})\right]\\
 & \leq\mathbf{E}\left[\sum_{i\in U}\ell\left(D_{i}\right)\right]\\
 & \overset{(a)}{=}\mathbf{E}\left[\sum_{i=1}^{n}\ell\left(D_{i}\right)\right]\\
 & \overset{(b)}{\leq}\mathbf{E}\left[n\ell\left(\frac{\widetilde{X}_{(n)}}{n}\right)\right]\\
 & \stackrel{(c)}{=}n\sum_{x=0}^{\infty}\mathbf{P}\left(\widetilde{X}_{(n)}>x\right)\left(\ell\left(\dfrac{x+1}{n}\right)-\ell\left(\dfrac{x}{n}\right)\right),
\end{align*}
where $(a)$ is because if $i\notin U,$ then $D_{i}=0$ and $\ell(D_{i})=0$.
For $(b),$ it follows by Jensen's inequality and $\sum_{i=1}^{n}D_{i}=X_{(n)}$.
For $(c)$, the equality follows by the fact that
\begin{align*}
 & \mathbf{E}\left[\ell\left(\frac{\widetilde{X}_{(n)}}{n}\right)\right]\\
 & =\mathbf{E}\left[\sum_{x=0}^{\widetilde{X}_{(n)}-1}\left(\ell\left(\frac{x+1}{n}\right)-\ell\left(\frac{x}{n}\right)\right)\right]\\
 & =\mathbf{E}\left[\sum_{x=0}^{\infty}\left(\ell\left(\frac{x+1}{n}\right)-\ell\left(\frac{x}{n}\right)\right)\textbf{1}\{\widetilde{X}_{(n)}>x\}\right]\\
 & =\sum_{x=0}^{\infty}\mathbf{P}\left(\widetilde{X}_{(n)}>x\right)\left(\ell\left(\frac{x+1}{n}\right)-\ell\left(\frac{x}{n}\right)\right).
\end{align*}
 Consider the term $\mathbf{P}\left(\widetilde{X}_{(n)}>x\right)$,
we have 
\begin{equation}
\mathbf{P}\left(\widetilde{X}_{(n)}>x\right)\leq\mathbf{P}\left(\stackrel[i=1]{n}{\cup}\left(\widetilde{X}_{i}>x\right)\right)\le cnx^{-\lambda}.\label{eq:thm1_prob_bound}
\end{equation}
 Hence, 
\begin{align*}
 & n\sum_{x=0}^{\infty}\mathbf{P}\left(\widetilde{X}_{(n)}>x\right)\left(\ell\left(\dfrac{x+1}{n}\right)-\ell\left(\dfrac{x}{n}\right)\right)\\
 & \leq n\sum_{x=0}^{\infty}\min(cnx^{-\lambda},1)\left(\ell\left(\dfrac{x+1}{n}\right)-\ell\left(\dfrac{x}{n}\right)\right).
\end{align*}
Letting $x_{0}=\left\lceil n^{1/\lambda}\right\rceil $, we have
\begin{align}
 & n\sum_{x=0}^{\infty}\min(cnx^{-\lambda},1)\left(\ell\left(\dfrac{x+1}{n}\right)-\ell\left(\dfrac{x}{n}\right)\right)\nonumber \\
 & \leq n\sum_{x=0}^{x_{0}-1}\left(\ell\left(\dfrac{x+1}{n}\right)-\ell\left(\dfrac{x}{n}\right)\right)\nonumber \\
 & \ \ \ \ +n\sum_{x=x_{0}}^{\infty}cnx^{-\lambda}\left(\ell\left(\dfrac{x+1}{n}\right)-\ell\left(\dfrac{x}{n}\right)\right).\label{eq:thm1_eq1}
\end{align}
Consider the term $n\sum_{x=0}^{x_{0}-1}\left(\ell((x+1)/n)-\ell(x/n)\right)$
in (\ref{eq:thm1_eq1}). Since it is a telescoping sum and $\ell(0)=0$,
we have 
\begin{align*}
 & n\sum_{x=0}^{x_{0}-1}\left(\ell\left(\dfrac{x+1}{n}\right)-\ell\left(\dfrac{x}{n}\right)\right)\\
 & =n\ell\left(\dfrac{x_{0}}{n}\right).
\end{align*}
Consider the term $n\sum_{x=x_{0}}^{\infty}cnx^{-\lambda}\left(\ell((x+1)/n)-\ell(x/n)\right)$
in (\ref{eq:thm1_eq1}). Since $\ell'(x)$ is non-increasing and $\ell'(x)>0$
when $x\geq0$, we have 
\[
\ell\left(\frac{x+1}{n}\right)-\ell\left(\frac{x}{n}\right)=\int_{\frac{x}{n}}^{\frac{x+1}{n}}\ell'(t)dt\leq\frac{\ell'(\frac{x}{n})}{n}.
\]
Therefore,
\begin{align*}
 & n\sum_{x=x_{0}}^{\infty}cnx^{-\lambda}\left(\ell\left(\dfrac{x+1}{n}\right)-\ell\left(\dfrac{x}{n}\right)\right)\\
 & \leq n\sum_{x=x_{0}}^{\infty}cnx^{-\lambda}\left(\frac{\ell'(\frac{x}{n})}{n}\right)\\
 & =\sum_{x=x_{0}}^{\infty}cnx^{-\lambda}\ell'\left(\frac{x}{n}\right).
\end{align*}
Hence,
\begin{align}
 & n\sum_{x=0}^{x_{0}-1}\left(\ell\left(\dfrac{x+1}{n}\right)-\ell\left(\dfrac{x}{n}\right)\right)\nonumber \\
 & +n\sum_{x=x_{0}}^{\infty}cnx^{-\lambda}\left(\ell\left(\dfrac{x+1}{n}\right)-\ell\left(\dfrac{x}{n}\right)\right)\nonumber \\
 & \leq n\ell\left(\dfrac{x_{0}}{n}\right)+\sum_{x=x_{0}}^{\infty}cnx^{-\lambda}\ell'\left(\frac{x}{n}\right)\nonumber \\
 & =n\ell\left(\dfrac{x_{0}}{n}\right)+\sum_{x=x_{0}}^{n}cnx^{-\lambda}\ell'\left(\frac{x}{n}\right)+\sum_{x=n+1}^{\infty}cnx^{-\lambda}\ell'\left(\frac{x}{n}\right)\nonumber \\
 & \stackrel{(d)}{\leq}n\ell\left(\dfrac{x_{0}}{n}\right)+cnx_{0}^{-\lambda}\ell'\left(\dfrac{x_{0}}{n}\right)+\int_{x_{0}}^{\infty}cnx^{-\lambda}\ell'\left(\dfrac{x}{n}\right)dx\nonumber \\
 & =2n\log\left(\dfrac{2x_{0}}{n}+1\right)+\dfrac{4cnx_{0}^{-\lambda}\log e}{\frac{2x_{0}}{n}+1}\nonumber \\
 & \ \ \ \ +\int_{x_{0}}^{n}\dfrac{4cnx^{-\lambda}\log e}{\frac{2x}{n}+1}dx+\int_{n}^{\infty}\dfrac{4cnx^{-\lambda}\log e}{\frac{2x}{n}+1}dx,\label{eq:thm1_eq2}
\end{align}
where $(d)$ follows by the fact that $\sum_{x=N}^{M}\ell(x)\leq\ell(N)+\int_{N}^{M}\ell(x)dx$
when $\ell(x)$ is a decreasing function.

Consider the term $2n\log\left(2x_{0}/n+1\right)$ in (\ref{eq:thm1_eq2}),
we have 

\begin{align*}
 & 2n\log\left(\dfrac{2x_{0}}{n}+1\right)\\
 & \leq2n\log\left(\frac{2n^{\frac{1}{\lambda}}+2}{n}+1\right)\\
 & \leq2n\left(\dfrac{2n^{\frac{1}{\lambda}}+2}{n}\right)\log e\\
 & =4n^{\frac{1}{\lambda}}\log e+4\log e\\
 & \leq8n^{\frac{1}{\lambda}}\log e.
\end{align*}
Consider the term $\int_{x_{0}}^{n}4cnx^{-\lambda}\log e/\left(\frac{2x}{n}+1\right)dx$
in (\ref{eq:thm1_eq2}). Since $\frac{2x_{0}}{n}+1>1$,
\begin{align*}
 & \int_{x_{0}}^{n}\dfrac{4cnx^{-\lambda}\log e}{\frac{2x}{n}+1}dx\\
 & \leq\int_{x_{0}}^{n}4cnx^{-\lambda}\log edx\\
 & \leq(4cn\log e)\frac{x_{0}^{1-\lambda}-n{}^{1-\lambda}}{\lambda-1}\\
 & \overset{(e)}{\leq}\frac{4c\left(n^{\frac{1}{\lambda}}+1\right)\log e-4cn^{2-\lambda}\log e}{\lambda-1}\\
 & \leq\dfrac{4cn^{\frac{1}{\lambda}}\log e}{\lambda-1},
\end{align*}
where $(e)$ follows by the fact that $n^{\frac{1}{\lambda}}\leq x_{0}\leq n{}^{\frac{1}{\lambda}}+1$
and $x_{0}^{-\lambda}\leq n{}^{-1}$.

Consider the term $\int_{n}^{\infty}4cnx^{-\lambda}\log e/\left(\frac{2x}{n}+1\right)dx$
in (\ref{eq:thm1_eq2}). Since $\frac{2x}{n}+1>\frac{2x}{n}$, we
have 
\begin{align*}
 & \int_{n}^{\infty}\dfrac{4cnx^{-\lambda}\log e}{\frac{2x}{n}+1}dx\\
 & \leq\int_{n}^{\infty}2cn^{2}x^{-\lambda-1}\log edx\\
 & =\frac{2cn^{2-\lambda}\log e}{\lambda}.
\end{align*}
Hence (\ref{eq:thm1_eq2}) can be bounded as, 
\begin{align*}
 & 2n\log\left(\dfrac{2x_{0}}{n}+1\right)+\dfrac{4cnx_{0}^{-\lambda}\log e}{\frac{2x_{0}}{n}+1}\\
 & \ \ \ \ +\int_{x_{0}}^{n}\dfrac{4cnx^{-\lambda}\log e}{\frac{2x}{n}+1}dx+\int_{n}^{\infty}\dfrac{4cnx^{-\lambda}\log e}{\frac{2x}{n}+1}dx,\\
 & \leq8n^{\frac{1}{\lambda}}\log e+4c\log e\\
 & \ \ \ \ +\dfrac{4cn^{\frac{1}{\lambda}}\log e}{\lambda-1}+\frac{2cn^{2-\lambda}\log e}{\lambda}\\
 & \leq\dfrac{26cn^{\frac{1}{\lambda}}}{\min(\lambda-1,1)}.
\end{align*}
Therefore, $\mathbf{E}\left[\sum_{i\in U}L(g(D_{i}))\right]\leq26cn^{1/\lambda}/\min(\lambda-1,1)$.

Consider $i\in V$. Note that $j_{i}=\min\{n\in\mathbb{N}:i+n\in U\}$
and $L(w_{i})=\left\lfloor 2\log j_{i}+1\right\rfloor +1\leq\ell(j_{i})$.
we have

\begin{align*}
\mathbf{E}\left[\sum_{i\in V}L(w_{i})\right] & \leq\mathbf{E}\left[\sum_{i\in V}\ell(j_{i})\right].
\end{align*}
Note that $\left|V\right|\leq\widetilde{X}_{(n)}$ and $\sum_{j_{i}:i\in V}j_{i}\leq n$.

\begin{align}
 & \mathbf{E}\left[\sum_{j_{i}:i\in V}\ell(j_{i})\right]\nonumber \\
 & =\mathbf{E}\left[\sum_{j_{i}:i\in V}\ell(j_{i})+\sum_{k=0}^{\widetilde{X}_{(n)}-\left|V\right|}\ell(0)\right]\nonumber \\
 & \overset{(f)}{\leq}\mathbf{E}\left[\widetilde{X}_{(n)}\ell\left(\dfrac{n}{\widetilde{X}_{(n)}}\right)\right]\nonumber \\
 & \overset{(g)}{\leq}\mathbf{E}\left(\widetilde{X}_{(n)}\right)\ell\left(\dfrac{n}{\mathbf{E}\left(\widetilde{X}_{(n)}\right)}\right),\label{eq:thm1_eq3}
\end{align}
where $(f)$ , $(g)$ follows by Jensen's inequality.

Since $x\ell(n/x)$ is an increasing function when $x\geq0$, (\ref{eq:thm1_eq3})
is bounded above when $\mathbf{E}(\widetilde{X}_{(n)})$ is bounded
above. For any non-negative integer-valued random variables $X$,
we have $\mathbf{E}\left[X\right]=\sum_{x=0}^{\infty}\mathbf{P}(X>x)$.
We have 
\begin{align*}
 & \mathbf{E}\left(\widetilde{X}_{(n)}\right)\\
 & =\sum_{x=0}^{\infty}\mathbf{P}\left(\widetilde{X}_{(n)}>x\right)\\
 & \leq\sum_{x=0}^{\infty}\min(cnx^{-\lambda},1)\\
 & =\sum_{x=0}^{x_{0}-1}1+\sum_{x=x_{0}}^{\infty}cnx^{-\lambda}\\
 & \leq x_{0}+cnx_{0}^{-\lambda}+\int_{x_{0}}^{\infty}cnx^{-\lambda}dx\\
 & \le n^{\frac{1}{\lambda}}+1+c+\frac{cn^{\frac{1}{\lambda}}}{\lambda-1}\\
 & \leq\dfrac{2c\lambda n^{\frac{1}{\lambda}}}{\lambda-1}+2c\\
 & \leq\frac{4c\lambda n^{\frac{1}{\lambda}}}{\lambda-1}.
\end{align*}
Hence,

\begin{align*}
 & \mathbf{E}\left(\widetilde{X}_{(n)}\right)\ell\left(\dfrac{n}{\mathbf{E}\left(\widetilde{X}_{(n)}\right)}\right)\\
 & \leq\frac{8c\lambda n^{\frac{1}{\lambda}}}{\lambda-1}\log\left(\dfrac{2n(\lambda-1)}{4\lambda cn^{\frac{1}{\lambda}}}+1\right)\\
 & \leq\frac{8c\lambda n^{\frac{1}{\lambda}}}{\lambda-1}\log\left(n^{1-\frac{1}{\lambda}}+1\right)\\
 & \leq\frac{8c\lambda n^{\frac{1}{\lambda}}}{\lambda-1}\log\left(2n^{1-\frac{1}{\lambda}}\right)\\
 & =\frac{8c\lambda n^{\frac{1}{\lambda}}}{\lambda-1}+16cn^{\frac{1}{\lambda}}\log\sqrt{n}\\
 & \leq\frac{24c\lambda n^{\frac{1}{\lambda}}\log(\sqrt{n}+1)}{\text{\ensuremath{\lambda-1}}}.
\end{align*}
Therefore, an upper bound for the $\mathbf{E}\left[\sum_{i=1}^{n}L(w_{i})\right]$
is 

\begin{align*}
 & \mathbf{E}\left[\sum_{i\in U}L(w_{i})\right]+\mathbf{E}\left[\sum_{i\in V}L(w_{i})\right]\\
 & \leq\dfrac{26cn^{\frac{1}{\lambda}}}{\min(\lambda-1,1)}+\frac{24c\lambda n^{\frac{1}{\lambda}}\log(\sqrt{n}+1)}{\text{\ensuremath{\lambda-1}}}\\
 & \leq\dfrac{50c\lambda n^{\frac{1}{\lambda}}\log(\sqrt{n}+1)}{\lambda-1}.
\end{align*}
\end{IEEEproof}
\begin{thm}
\label{thm:p_int-2}Let $P$ be a distribution over positive integers.
If $P$ satisfies the bound $P(X>x)\leq ce^{-\lambda x}$ for all
integer $x\ge0$, where $c>1$ and $\lambda>0$, then the expected
codeword length of difference run-length coding scheme for $P$ is
upper bounded as
\begin{align*}
\mathbf{E}(L(W)) & \leq\dfrac{13(2\lambda+1)c}{\lambda}\log^{2}\left(n+1\right).
\end{align*}
\end{thm}
\begin{IEEEproof}
By replacing 
\[
\mathbf{P}\left(\widetilde{X}_{(n)}>x\right)\leq\mathbf{P}\left(\stackrel[i=1]{n}{\cup}\left(\widetilde{X}_{i}>x\right)\right)\le cnx^{-\lambda}
\]

in (\ref{eq:thm1_prob_bound}) with 
\[
\mathbf{P}\left(\widetilde{X}_{(n)}>x\right)\leq\mathbf{P}\left(\stackrel[i=1]{n}{\cup}\left(\widetilde{X}_{i}>x\right)\right)\le cne^{-\lambda x},
\]

and replacing $x_{0}=\left\lceil n^{1/\lambda}\right\rceil $ in (\ref{eq:thm1_eq1})
with $x_{0}=\left\lceil (\ln n)/\lambda\right\rceil $, we can rewrite
(\ref{eq:thm1_eq2}) to

\begin{align}
 & 2n\log\left(\dfrac{2x_{0}}{n}+1\right)+\dfrac{4cne^{-\lambda x_{0}}\log e}{\frac{2x_{0}}{n}+1}\nonumber \\
 & +\int_{x_{0}}^{\infty}\dfrac{4cne^{-\lambda x}\log e}{\frac{2x}{n}+1}dx.\label{eq:thm5-eq1}
\end{align}

Consider the term $2n\log\left(\dfrac{2x_{0}}{n}+1\right)$ in (\ref{eq:thm5-eq1}),
we have
\begin{align*}
 & 2n\log\left(\dfrac{2x_{0}}{n}+1\right)\\
 & \leq2n\log\left(\frac{2\left(\frac{\ln n}{\lambda}+1\right)}{n}+1\right)\\
 & \leq2n\left(\dfrac{2\left(\frac{\ln n}{\lambda}+1\right)}{n}\right)\log e\\
 & =4\left(\frac{(\ln2)\log n}{\lambda}+1\right)\log e.
\end{align*}

Consider the term $4cne^{-\lambda x_{0}}\log e/(2x_{0}/n+1)$ in (\ref{eq:thm5-eq1}).
Since $2x_{0}/n+1\geq1$ and $ne^{-\lambda x_{0}}\leq1$, 
\begin{align*}
 & \dfrac{4cne^{-\lambda x_{0}}\log e}{\frac{2x_{0}}{n}+1}\\
 & \leq4c\log e.
\end{align*}

Consider the term $\int_{x_{0}}^{\infty}4cne^{-\lambda x}\log e/(2x/n+1)dx$
in (\ref{eq:thm5-eq1}). Since $2x_{0}/n+1\geq1$,
\begin{align*}
 & \int_{x_{0}}^{\infty}\dfrac{4cne^{-\lambda x}\log e}{\frac{2x}{n}+1}dx\\
 & \leq\int_{x_{0}}^{\infty}4cne^{-\lambda x}\log edx\\
 & \leq(4cn\log e)\frac{e^{-\lambda x_{0}}}{\lambda}\\
 & \overset{(a)}{\leq}\dfrac{(4cn\log e)n^{-1}}{\lambda}\\
 & \leq\dfrac{4c\log e}{\lambda},
\end{align*}

where $(a)$ follows by the fact that $(\ln n)/\lambda\leq x_{0}\leq(\ln n)/\lambda+1$
and $e^{-\lambda x_{0}}\leq n^{-1}$.

Hence (\ref{eq:thm5-eq1}) can be bounded as, 
\begin{align*}
 & 2n\log\left(\dfrac{2x_{0}}{n}+1\right)+\dfrac{4cnx_{0}^{-\lambda}\log e}{\frac{2x_{0}}{n}+1}\\
 & \ \ \ \ +\int_{x_{0}}^{\infty}\dfrac{4cnx^{-\lambda}\log e}{\frac{2x}{n}+1}dx,\\
 & \leq4\left(\frac{(\ln2)\log n}{\lambda}+1\right)\log e+4c\log e\\
 & \ \ \ \ +\dfrac{4c\log e}{\lambda}\\
 & \leq\dfrac{4\log n}{\lambda}+6c\left(2+\dfrac{1}{\lambda}\right).
\end{align*}

Therefore, $\mathbf{E}\left[\sum_{i\in U}L(w_{i}))\right]\leq\dfrac{4\log n}{\lambda}+6c\left(2+\dfrac{1}{\lambda}\right)$.

For finding an upper bound of $\mathbf{E}\left[\sum_{i\in V}L(w_{i})\right]$,
we use same argument in Theorem (\ref{thm:p_int}). We start with
finding an upper bound for $\mathbf{E}\left(\widetilde{X}_{(n)}\right).$

\begin{align*}
 & \mathbf{E}\left(\widetilde{X}_{(n)}\right)\\
 & =\sum_{x=0}^{\infty}\mathbf{P}\left(\widetilde{X}_{(n)}>x\right)\\
 & \leq\sum_{x=0}^{\infty}\min(cne^{-\lambda x},1)\\
 & =\sum_{x=0}^{x_{0}-1}1+\sum_{x=x_{0}}^{\infty}cne^{-\lambda x}\\
 & \leq x_{0}+cne^{-\lambda x_{0}}+\int_{x_{0}}^{\infty}cne^{-\lambda x}dx\\
 & \leq\dfrac{\ln n}{\lambda}+1+c+\dfrac{c}{\lambda}\\
 & \leq\dfrac{(\ln2)\log n}{\lambda}+(2+\frac{1}{\lambda})c.
\end{align*}

Hence,

\begin{align*}
 & \mathbf{E}\left(\widetilde{X}_{(n)}\right)\ell\left(\dfrac{n}{\mathbf{E}\left(\widetilde{X}_{(n)}\right)}\right)\\
 & \leq2\left(\dfrac{(\ln2)\log n}{\lambda}+(2+\frac{1}{\lambda})c\right)\log\left(\dfrac{2n\lambda}{(\ln2)\log n+(2\lambda+1)c}+1\right)\\
 & \leq2\left(\dfrac{(\ln2)\log n}{\lambda}+(2+\frac{1}{\lambda})c\right)\log\left(n+1\right)\\
 & \leq\left(\dfrac{2\ln2+(2\lambda+1)c}{\lambda}\right)\log^{2}\left(n+1\right).
\end{align*}
\\
Therefore, an upper bound for the $\mathbf{E}\left[\sum_{i=1}^{n}L(w_{i})\right]$
is 

\begin{align*}
 & \mathbf{E}\left[\sum_{i\in U}L(w_{i})\right]+\mathbf{E}\left[\sum_{i\in V}L(w_{i})\right]\\
 & \leq\dfrac{4\log n}{\lambda}+6c\left(2+\dfrac{1}{\lambda}\right)+\left(\dfrac{2\ln2+(2\lambda+1)c}{\lambda}\right)\log^{2}\left(n+1\right)\\
 & \leq\dfrac{4+6(2\lambda+1)c+2\ln2+(2\lambda+1)c}{\lambda}\log^{2}\left(n+1\right)\\
 & \leq\dfrac{13(2\lambda+1)c}{\lambda}\log^{2}\left(n+1\right).
\end{align*}
\end{IEEEproof}
This encoding scheme also plays an essential role in establishing
a coding scheme for the case where $P$ is a continuous distribution
over $[0,\infty)$ with a non-increasing pdf, which will be discussed
in Section \ref{sec:p_inf}.

\section{$P$ is a continuous distribution over $[0,1]$ with a non-increasing
pdf \label{sec:p_01}}

We develop another coding scheme for the case where $P$ is a continuous
distribution over $[0,1]$ with a non-increasing pdf, which is another
building block for the case where $P$ is a distribution over $[0,\infty)$
in Section \ref{sec:p_inf}. Our scheme is based on the dyadic decomposition
construction in \cite{li2017distributed,li2018universal}. 
\begin{defn}
Let $f$ be the pdf of the distribution $P$, which is a non-increasing
function over $[0,\infty)$. For $k\in\mathbb{Z}_{\geq0}$ and $a\in[0:\max(2^{k-1}-1,0)]$,
define the rectangle 

\begin{align*}
 & R(k,a)\\
 & =\left[2^{-k+1}a,2^{-k}(2a+1)\right)\\
 & \;\;\;\;\;\times\left[f\left(2^{-k+1}(a+1)\right),f\left(2^{-k}(2a+1)\right)\right)\\
 & \subseteq\mathbb{R}^{2}.
\end{align*}
Consider the positive part of the hypograph of $f$ defined as $\mathrm{hyp}f^{+}=\{(x,y):x\in\mathbb{R}_{\geq0},0\leq y\leq f(x)\}$.
Note that $\{R(k,a)\}$ is a partition of $\mathrm{hyp}f^{+}$ (except
a set of measure zero) into rectangles. Every point $x$ in the interior
of $\mathrm{hyp}f^{+}$ is contained in only one rectangle $R(k,a)$. 

Note that $k,a$ can be 0. When encoding $k,a$, we will use shifted
Elias gamma code defined as follows. Let $g_{s}(x)=g(x+1)$, where
$g$ is the Elias gamma encoding function.
\end{defn}
We are now ready to define our coding scheme for the continuous distribution
$P$ over $[0,1]$ with a non-increasing pdfs.
\begin{defn}
\label{def:scheme_p_01}The coding scheme for the case where $P$
is a continuous distribution over $[0,1]$ with a non-increasing pdf
consists of:
\begin{enumerate}
\item Encoder: 
\begin{enumerate}
\item After observing $P$, the encoder generates $n$ i.i.d. points $p_{1},p_{2},...,p_{n}\in\mathrm{hyp}f^{+}$
uniformly over $\mathrm{hyp}f^{+}$. 
\item Let $U=\{(k,a):R(k,a)\cap\{p_{1},p_{2},\ldots,p_{n}\}\neq\emptyset\}$.
Assume $U=\{(k_{1},a_{1}),(k_{2},a_{2}),\ldots(k_{|U|},a_{|U|})\}$,
where $(k_{i},a_{i})$ are ordered in lexicographic order.
\item Let $w_{i}=g_{s}(k_{i})||g_{s}(a_{i})||g(N_{i})$ for $i\in[1:|U|]$,
where $N_{i}=|R(k_{i},a_{i})\cap\{p_{1},p_{2},\ldots,p_{n}\}|$. The
encoder then sends the codeword $W$, which is the concatenation of
$w_{i}$, i.e., $W=w_{1}||w_{2}||\cdots||w_{\left|U\right|}$, to
the decoder.
\end{enumerate}
\item Decoder:
\begin{enumerate}
\item Upon receiving $W$, the decoder recovers $k_{i},a_{i},N_{i}$ for
$i\in[1:|U|]$.
\item For each $i$, the decoder generates $N_{i}$ points uniformly over
$R(k_{i},a_{i})$. It collects all the $x$-coordinate of the generated
points, shuffle these numbers uniformly at random and outputs the
shuffled sequence as $X_{1},X_{2},\ldots,X_{n}$.
\end{enumerate}
\end{enumerate}
\end{defn}
We present the following theorem which shows that the codeword length
grows sub-linearly in $n$ when $P$ follows a non-increasing pdf. 
\begin{thm}
\label{thm:p_01}The expected codeword length of the above coding
scheme for the case where $P$ is a distribution over $[0,1]$ with
a non-increasing pdf is
\[
\mathbf{E}(L(W))\leq92\sqrt{nf(0)}(\log(\sqrt{nf(0)}+1)).
\]
\end{thm}
\begin{example}
\label{exa:tri}Consider the following pdf $f$ over $[0,1]$,
\[
f(x)=\begin{cases}
2-2x & ,\mbox{ if \ensuremath{0\leq x\leq1}}\\
0 & ,\mbox{ otherwise}.
\end{cases}
\]
Figure \ref{fig:tri_decompose} depicts the decomposition of this
pdf into rectangles. Figure \ref{fig:loglog} depicts a log-log plot
of the expected codeword length (computed by listing all rectangles
with width at least $2^{-8}$) versus $n$, compared to the bound
in Theorem \ref{thm:p_01}. Notice that the growth rate of the expected
codeword length has a similar order as our bound, and they are both
sublinear (which can be observed from their slopes in the log-log
plot which are less than $1$).

\begin{figure}[h]
\begin{raggedright}
\includegraphics[scale=0.19]{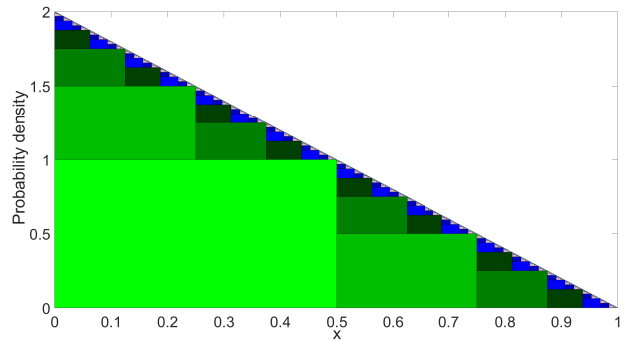}
\par\end{raggedright}
\caption{\label{fig:tri_decompose}Decomposition of the distribution in Example
\ref{exa:tri}.}
\end{figure}

\begin{figure}[h]
\begin{raggedright}
\includegraphics[viewport=200bp 0bp 1440bp 704bp,scale=0.22]{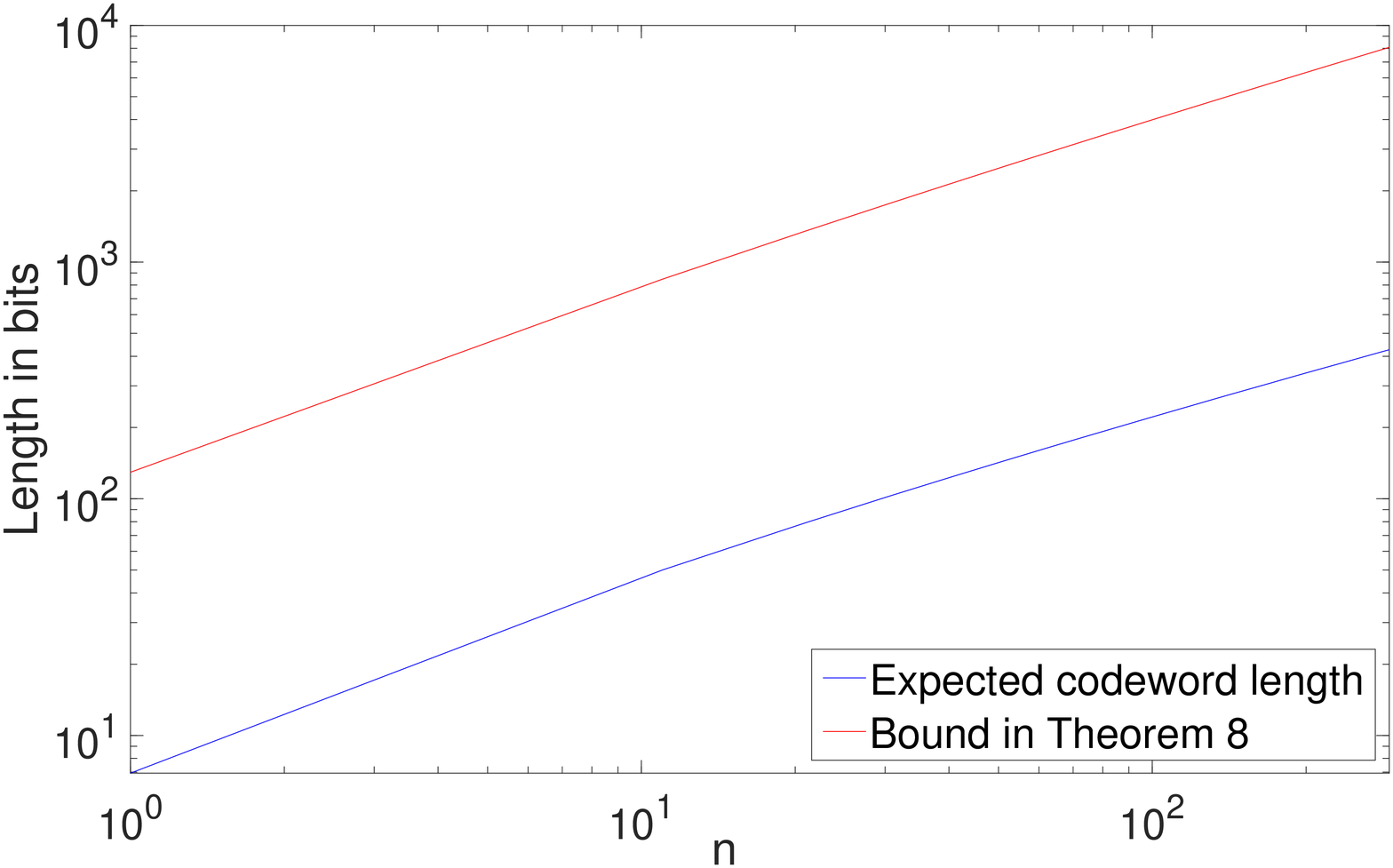}
\par\end{raggedright}
\caption{\label{fig:loglog}Log-log plot of the expected codeword length and
the bound in Theorem \ref{thm:p_01} for Example \ref{exa:tri}. }

\end{figure}

We now prove Theorem \ref{thm:p_01}.
\end{example}
\begin{IEEEproof}
Let $N_{k,a}=|R(k,a)\cap\{p_{1},p_{2},\ldots,p_{n}\}|$. Since $n$
points are generated randomly and independently on the $\mathrm{hyp}f^{+}$,
$N_{k,a}$ is a random variable and follows a distribution $\mathrm{Binomial}(n,A(k,a))$,
where $A(k,a)$ is the area of rectangle $R(k,a)$. More specifically,
$A(k,a)=2^{-k}\left(f\left(2^{-k}(2a+1)\right)-f\left(2^{-k+1}(a+1)\right)\right)$.
Also, the probability to include the triple $(k,a,N_{k,a})$ in the
encoding $W$ is equal to $\mathbf{P}(N_{k,a}\geq1)$. The expected
codeword length $\mathbf{E}\left[L(W)\right]$ can be calculated by
summing the expected codeword length for all possible triples. 

\begin{align}
 & \mathbf{E}\left[L(W)\right]\nonumber \\
 & =\sum_{k=0}^{\infty}\sum_{a=0}^{\max(2^{k-1}-1,0)}\left(L\left(g_{s}(k)\right)+L\left(g_{s}(a)\right)\right)\mathbf{P}(N_{k,a}\geq1)\nonumber \\
 & \ \ \ \ +\sum_{k=0}^{\infty}\sum_{a=0}^{\max(2^{k-1}-1,0)}\mathbf{E}\left[L\left(g(N_{k,a})\right)\right]\nonumber \\
 & =\left(L\left(g_{s}(0)\right)+L\left(g_{s}(0)\right)\right)\mathbf{P}(N_{0,0}\geq1)\nonumber \\
 & \ \ \ \ +\sum_{k=1}^{\infty}\sum_{a=0}^{2^{k-1}-1}\left(L\left(g_{s}(k)\right)\right)\left(1-\left(1-A(k,a)\right){}^{n}\right)\nonumber \\
 & \ \ \ \ +\sum_{k=1}^{\infty}\sum_{a=0}^{2^{k-1}-1}\left(L\left(g_{s}(a)\right)\right)\left(1-\left(1-A(k,a)\right){}^{n}\right)\nonumber \\
 & \ \ \ \ +\mathbf{E}\left[L(g(N_{0,0}))\right]\nonumber \\
 & \ \ \ \ +\sum_{k=1}^{\infty}\sum_{a=0}^{2^{k-1}-1}\mathbf{E}\left[L\left(g(N_{k,a})\right)\right].\label{eq:thm2_eq1}
\end{align}
Note that $\left(L\left(g_{s}(0)\right)+L\left(g_{s}(0)\right)\right)\mathbf{P}(N_{0,0}\geq1)\leq(1+1)(1)=2$.

Consider the term $1-\left(1-A(k,a)\right){}^{n}$ in (\ref{eq:thm2_eq1}).
Let $k_{0}=\left\lfloor \log\left(\sqrt{nf(0)}+1\right)\right\rfloor $.
If $k\leq k_{0}$, then bound $1-(1-A(k,a))^{n}$ above by 1. Otherwise,
bound $1-(1-A(k,a))^{n}$ above by $nA(k,a)$. Consider the term $\sum_{k=1}^{\infty}\sum_{a=0}^{2^{k-1}-1}\left(L\left(g_{s}(k)\right)+L\left(g_{s}(a)\right)\right)\left(1-\left(1-A(k,a)\right){}^{n}\right),$
we have
\begin{align*}
 & \sum_{k=1}^{\infty}\sum_{a=0}^{2^{k-1}-1}\left(L\left(g_{s}(k)\right)+L\left(g_{s}(a)\right)\right)\left(1-\left(1-A(k,a)\right){}^{n}\right)\\
 & \leq\sum_{k=1}^{k_{0}}\sum_{a=0}^{2^{k-1}-1}\left(L\left(g_{s}(k)\right)+L\left(g_{s}(a)\right)\right)\\
 & \ \ \ \ +\sum_{k=k_{0}+1}^{\infty}\sum_{a=0}^{2^{k-1}-1}\left(L\left(g_{s}(k)\right)+L\left(g_{s}(a)\right)\right)nA(k,a).
\end{align*}
Consider the term $\sum_{k=1}^{k_{0}}\sum_{a=0}^{2^{k-1}-1}L\left(g_{s}(k)\right)$
in (\ref{eq:thm2_eq1}). We have 
\begin{align*}
 & \sum_{k=1}^{k_{0}}\sum_{a=0}^{2^{k-1}-1}L\left(g_{s}(k)\right)\\
 & \leq\sum_{k=1}^{k_{0}}2^{k}\log(2k+3)\\
 & \stackrel{(a)}{\leq}\sum_{k=1}^{k_{0}}(k+1)(2^{k+1})\\
 & =2^{k_{0}+2}k_{0}\\
 & \leq4(\sqrt{nf(0)}+1)\log(\sqrt{nf(0)}+1)\\
 & \leq8\sqrt{nf(0)}\log(\sqrt{nf(0)}+1),
\end{align*}
where $(a)$ follows from the fact that $\log(2k+3)\leq2k+2$ when
$k\geq1$. 

Consider the term $\sum_{k=1}^{k_{0}}\sum_{a=0}^{2^{k-1}-1}L\left(g_{s}(a)\right)$
in (\ref{eq:thm2_eq1}). We have 
\begin{align*}
 & \sum_{k=1}^{k_{0}}\sum_{a=0}^{2^{k-1}-1}L\left(g_{s}(a)\right)\\
 & \leq\sum_{k=1}^{k_{0}}\sum_{a=0}^{2^{k-1}-1}2\log(2a+3)\\
 & \leq\sum_{k=1}^{k_{0}}\int_{0}^{2^{k-1}}2\log(2a+3)da\\
 & =\sum_{k=1}^{k_{0}}(2^{k}+3)\log(2^{k}+3)-3\log3-2^{k}\\
 & \leq\sum_{k=1}^{k_{0}}2^{k}\log(2^{k+2})\\
 & =\sum_{k=1}^{k_{0}}(k+2)2^{k}\\
 & =2(2^{k_{0}}k_{0}+2^{k_{0}}-1)\\
 & \leq4(\sqrt{nf(0)}+1)\log(\sqrt{nf(0)}+1)\\
 & \leq8\sqrt{nf(0)}\log(\sqrt{nf(0)}+1).
\end{align*}
Consider the term $\sum_{k=k_{0}+1}^{\infty}\sum_{a=0}^{2^{k-1}-1}g_{s}(k)nA(k,a)$
in (\ref{eq:thm2_eq1}). We have
\begin{align*}
 & \sum_{k=k_{0}+1}^{\infty}\sum_{a=0}^{2^{k-1}-1}g_{s}(k)nA(k,a)\\
 & \leq n\sum_{k=k_{0}+1}^{\infty}\sum_{a=0}^{2^{k-1}-1}\log(2k+3)2^{-k+1}\left[f\left(2^{-k}(2a+1)\right)\right.\\
 & \ \ \ \ \left.-f\left(2^{-k+1}(a+1)\right)\right]\\
 & \stackrel{(b)}{\leq}n\sum_{k=k_{0}+1}^{\infty}\log(2k+3)2^{-k+1}f(0)\\
 & \overset{(c)}{\leq}nf(0)\sum_{k=k_{0}+1}^{\infty}(k+1)2^{-k+1}\\
 & =nf(0)2^{-k_{0}+1}(k_{0}+3)\\
 & \leq nf(0)2^{-k_{0}+3}k_{0}\\
 & \leq nf(0)2^{4-\log\left(\sqrt{nf(0)}+1\right)}\log\left(\sqrt{nf(0)}+1\right)\\
 & =\dfrac{16nf(0)}{\sqrt{nf(0)}+1}\log(\sqrt{nf(0)}+1)\\
 & \leq16\sqrt{nf(0)}\log(\sqrt{nf(0)}+1),
\end{align*}
where $(b)$ follow by the fact that $\sum_{a=0}^{2^{k-1}-1}f\left(2^{-k}(2a+1)\right)-f\left(2^{-k+1}(a+1)\right)\leq f(2^{-k})\leq f(0).$
For $(c)$, $k+1\geq\log(2k+3)$, when $k\geq2$.

Consider the term $\sum_{k=k_{0}+1}^{\infty}\sum_{a=0}^{2^{k-1}-1}g_{s}(a)nA(k,a)$
in (\ref{eq:thm2_eq1}). We have\\
\begin{align*}
 & \sum_{k=k_{0}+1}^{\infty}\sum_{a=0}^{2^{k-1}-1}g_{s}(a)nA(k,a)\\
 & =n\sum_{k=k_{0}+1}^{\infty}\sum_{a=0}^{2^{k-1}-1}\log(2a+3)2^{1-k}\Big(f\left(2^{-k}(2a+1)\right)\\
 & \ \ \ \ -f\left(2^{-k+1}(a+1)\right)\Big)\\
 & \stackrel{(d)}{\leq}n\sum_{k=k_{0}+1}^{\infty}\sum_{a=0}^{2^{k-1}-1}\log(2^{k+1})2^{1-k}\Big(f\left(2^{-k}(2a+1\right)\\
 & \ \ \ \ -f\left(2^{-k+1}(a+1)\right)\Big)\\
 & \leq nf(0)\sum_{k=k_{0}+1}^{\infty}\left(k+1\right)2^{1-k}\\
 & \leq16\sqrt{nf(0)}\log(\sqrt{nf(0)}+1),
\end{align*}
where $(d)$ follows from the fact that $\log(2a+3)\leq\log(2^{k}+1)\leq\log(2^{k+1})$
for $a$$\in\left[0:2^{k-1}-1\right]$.

Consider the term $\mathbf{E}\left[L(g(N_{0,0}))\right]$ in (\ref{eq:thm2_eq1}).
We have

\begin{align*}
 & \mathbf{E}\left[L(g(N_{0,0}))\right]\\
 & \leq\mathbf{E}(2\log(2N_{0,0}+1))\\
 & \overset{(e)}{\leq}2\log(2\mathbf{E}(N_{0,0})+1)\\
 & =2\log(2nf(1)+1)\\
 & \leq2\log(3nf(0))\\
 & \leq4+4\log(\sqrt{nf(0)}+1),
\end{align*}
where $(e)$ follows from Jensen's inequality.

Consider the term $\sum_{k=1}^{\infty}\sum_{a=0}^{2^{k-1}-1}\mathbf{E}\left[L\left(g(N_{k,a})\right)\right]$
in (\ref{eq:thm2_eq1}). We have
\begin{align*}
 & \sum_{k=1}^{\infty}\sum_{a=0}^{2^{k-1}-1}\mathbf{E}\left[L\left(g(N_{k,a})\right)\right]\\
 & \leq\sum_{k=1}^{\infty}\sum_{a=0}^{2^{k-1}-1}\mathbf{E}\left[2\log\left(2N_{k,a}+1\right)\right]\\
 & \stackrel{(f)}{\leq}\sum_{k=1}^{\infty}\sum_{a=0}^{2^{k-1}-1}2\log(2\mathbf{E}(N_{k,a})+1)\\
 & =\sum_{k=1}^{\infty}\sum_{a=0}^{2^{k-1}-1}2\log(2nA(k,a)+1)\\
 & =\sum_{k=1}^{\infty}\sum_{a=0}^{2^{k-1}-1}2\log\Big(\\
 & \ \ \ \ 2n\left(2^{-k}\left(f\left(2^{-k}(2a+1)\right)-f\left(2^{-k+1}(a+1)\right)\right)\right)+1\Big)\\
 & \overset{(g)}{\leq}\sum_{k=1}^{\infty}2^{k}\log\bigg(\\
 & \ \ \ \ \dfrac{4n}{2^{2k}}\sum_{a=0}^{2^{k-1}-1}\left[f\left(2^{-k}(2a+1)\right)-f\left(2^{-k+1}(a+1)\right)\right]+1\bigg)\\
 & \leq\sum_{k=1}^{\infty}2^{k}\log\left(\dfrac{4nf(0)}{2^{2k}}+1\right),
\end{align*}
where $(f)$and $(g)$ follows by Jensen's inequality. 

Let $k_{1}=\left\lfloor \log(\sqrt{4nf(0)}\right\rfloor $. We have\\
\begin{align}
 & \sum_{k=1}^{\infty}2^{k}\log\left(\dfrac{4nf(0)}{2^{2k}}+1\right)\nonumber \\
 & =\sum_{k=1}^{k_{1}}2^{k}\log\left(\dfrac{4nf(0)}{2^{2k}}+1\right)\nonumber \\
 & \ \ \ \ +\sum_{k=k_{1}+1}^{\infty}2^{k}\log\left(\dfrac{4nf(0)}{2^{2k}}+1\right)\nonumber \\
 & \overset{(h)}{\leq}\sum_{k=1}^{k_{1}}2^{k}\log\left(\dfrac{8nf(0)}{2^{2k}}\right)+\int_{k_{1}}^{\infty}2^{k}\log\left(\dfrac{4nf(0)}{2^{2k}}+1\right)dk,\label{eq:thm2_eq2}
\end{align}
where $(h)$ follows from the fact that $\dfrac{4nf(0)}{2^{2k}}\geq1$
when $k\leq k_{1}$ and $2^{k}\log\left(\dfrac{4nf(0)}{2^{2k}}+1\right)$
is decreasing when $k\geq k_{1}+1$.

Consider the term $\sum_{k=1}^{k_{1}}2^{k}\log\left(\dfrac{8nf(0)}{2^{2k}}\right)$
in (\ref{eq:thm2_eq2}),\\
\begin{align*}
 & \sum_{k=1}^{k_{1}}2^{k}\log\left(\dfrac{8nf(0)}{2^{2k}}\right)\\
 & =2(2^{k_{1}}-1)\log(8nf(0))-4(2^{k_{1}}k_{1}-2^{k_{1}}+1)\\
 & \leq2^{k_{1}+1}\log(8nf(0))-4(2^{k_{1}}k_{1}-2^{k_{1}})\\
 & \overset{(i)}{\leq}20\sqrt{nf(0)}+4\sqrt{nf(0)}\log(\sqrt{nf(0)})\\
 & \leq24\sqrt{nf(0)}\log(\sqrt{nf(0)}+1),
\end{align*}
where $(i)$ follows from the fact that $\log(\sqrt{4nf(0)})-1\leq k_{1}\leq\log(\sqrt{4nf(0)}$,
and thus $\sqrt{nf(0)}\leq2^{k_{1}}\leq2\sqrt{nf(0)}.$

Consider the term $\int_{k_{1}}^{\infty}2^{k}\log\left(\dfrac{4nf(0)}{2^{2k}}+1\right)dk$
in (\ref{eq:thm2_eq2}),

\begin{align*}
 & \int_{k_{1}}^{\infty}2^{k}\log\left(\dfrac{4nf(0)}{2^{2k}}+1\right)dk\\
 & =2\log^{2}e\sqrt{4nf(0)}\arctan\left(\frac{\sqrt{4nf(0)}}{2^{k_{1}}}\right)\\
 & \ \ \ \ -2^{k_{1}}\log\left(\frac{2^{2k_{1}}+4nf(0)}{2^{2k_{1}}}\right)\log e\\
 & \leq4\sqrt{nf(0)}\arctan(2)\log^{2}e\\
 & \leq10\sqrt{nf(0)}.
\end{align*}
Therefore,

\begin{align*}
 & \mathbf{E}\left[L(W)\right]\\
 & \leq72\sqrt{nf(0)}(\log(\sqrt{nf(0)}+1))+10\sqrt{nf(0)}\\
 & \ \ \ \ +4+4\log(\sqrt{nf(0)}+1)+2\\
 & \leq92\sqrt{nf(0)}(\log(\sqrt{nf(0)}+1)).
\end{align*}
\end{IEEEproof}

\section{$P$ is a distribution over $[0,\infty)$ with a non-increasing pdf\label{sec:p_inf}}

With the previous two coding schemes as our building blocks, we can
develop a coding scheme for the case where $P$ is a continuous distribution
over $\left[0,\infty\right)$ with a non-increasing pdf $f$ based
on the dyadic decomposition construction in \cite{li2017distributed,li2018universal}.
If $P$ satisfy a power law bound, we show the growth rate of expected
codeword length is sub-linear. 
\begin{defn}
\label{def:The-coding-scheme, =00005B0,inf)}The coding scheme for
the case where $P$ is a distribution over $[0,\infty)$ with a non-increasing
pdf consists of:
\begin{enumerate}
\item Encoder: 
\begin{enumerate}
\item After the encoder observes $P$, it generates i.i.d. $\widetilde{X}_{1},\widetilde{X}_{2},...,\widetilde{X}_{n}\sim P$.
\item Apply the difference run-length encoding scheme in Definition \ref{def:scheme_p_int}
to encode $\lceil\widetilde{X}_{1}\rceil,\ldots,\lceil\widetilde{X}_{n}\rceil$.
Let its output be $W_{\mathrm{int}}\in\{0,1\}^{*}$.
\item Let $n_{i}=|\{j:\widetilde{X}_{j}\in[i-1,i)\}|$. For each positive
integer $i$ where $n_{i}>0$, apply the scheme in Definition \ref{def:scheme_p_01}
to generate $n_{i}$ points with pdf 
\[
f_{i}(x)=\frac{f(x+i-1)}{\int_{i-1}^{i}f(t)dt}.
\]
Let its output be $W_{i}\in\{0,1\}^{*}$. For $i$ where $n_{i}=0$,
let $W_{i}=\emptyset$.
\item The encoder outputs $W=W_{\mathrm{int}}\Vert W_{1}\Vert W_{2}\Vert\cdots$.
\end{enumerate}
\item Decoder:
\begin{enumerate}
\item Upon receiving $W$, the decoder decodes $W_{\mathrm{int}}$ and recovers
the multiset $\{\lceil\widetilde{X}_{1}\rceil,\ldots,\lceil\widetilde{X}_{n}\rceil\}$,
and hence recovers $n_{i}$ for nonnegative integers $i$.
\item For each $i$ where $n_{i}>0$, the decoder decodes $W_{i}$ using
$W$, and use the decoding scheme in Definition \ref{def:scheme_p_01}
to generate i.i.d. $X_{i,1},\ldots,X_{i,n_{i}}\sim f_{i}$.
\item The decoder randomly shuffles $\{X_{i,j}\}_{i\ge1,\,j\in[1:n_{i}]}$
and output the shuffled sequence.
\end{enumerate}
\end{enumerate}
\end{defn}
We present the following theorem which shows that the codeword length
grows sub-linearly in $n$ when $P$ follows a non-increasing pdf
and satisfies a power law bound. 
\begin{thm}
\label{thm:p_inf}The expected codeword length of the above coding
scheme, for the case where $P$ is a distribution over $[0,\infty)$
with a non-increasing pdf, and $P$ satisfies the bound $\text{\ensuremath{\mathbf{P}}}(X>x)\leq cx^{-\lambda}$
for all $x\in[0,\infty)$, where $c>1$ and $\lambda>1$, is bounded
above as

\begin{align*}
\mathbf{E}\left(L\left(W\right)\right) & \leq\dfrac{418c(\lambda+1)\max(n^{1/\lambda},\sqrt{n})\max(\sqrt{f(0)},1)}{\min(\lambda-1,1)}\\
 & \ \ \ \ \cdot\log(\sqrt{n\max(f(0),1)}+1).
\end{align*}
\end{thm}
Before proving the theorem, we review the concept of majorization
for non-increasing functions \cite{chong1974some}.
\begin{defn}
[Majorization] Let $f,g$ be two continuous non-increasing functions
over $[0,\infty)$. It is said that $f$ is majorized by $g$, denoted
by $f\prec g$, if $\int_{0}^{x}f(t)dt\leq\int_{0}^{x}g(t)dt$ for
any $x\ge0$.
\end{defn}
We state the following equivalent characterization of majorization,
which is proved in \cite[Theorem 2.5]{chong1974some}.
\begin{lem}
[\cite{chong1974some}]\label{lem:maj_concave}Let $f,g$ be two
continuous functions. $f\prec g$ if and only if 
\begin{align*}
\int_{0}^{\infty}\phi(f(t))dt & \geq\int_{0}^{\infty}\phi(g(t))dt
\end{align*}

for all concave function $\phi:\mathbb{R\rightarrow\mathbb{R}}.$
\end{lem}
Before we prove Theorem \ref{thm:p_inf}, we show the following lemma.
\begin{lem}
\label{lem:power_bd}Let $f$ be a non-increasing pdf over $[0,\infty)$
that satisfies the bound $\int_{x}^{\infty}f(t)dt\leq cx^{-\lambda}$
for any $x\in[0,\infty)$. Let $f^{*}:[0,\infty)\rightarrow\mathbb{R}$
be a pdf defined as

\begin{align*}
f^{*}(x)= & \begin{cases}
c\lambda t_{0}^{-\lambda-1} & ,x\leq t_{0}\\
c\lambda x^{-\lambda-1} & ,x>t_{0},
\end{cases}
\end{align*}
where $t_{0}=\left(c(\lambda+1)\right)^{1/\lambda}$. Then $f\succ f^{*}.$
\end{lem}
\begin{IEEEproof}
Note that $\int_{0}^{\infty}f^{*}(x)dx=1.$ Further note that $f(0)\geq f^{*}(0),$otherwise
$\int_{0}^{\infty}f(x)dx<\int_{0}^{\infty}f^{*}(x)dx=1.$ Suppose
$\int_{0}^{a}f(x)dx<\int_{0}^{a}f^{*}(x)dx$ for some $a$. Suppose
$a\leq t_{0}.$ Note that $f(a)<f^{*}(a)$ and thus $f(x)<f^{*}(x)$
when $x\geq a$. Thus, $\int_{0}^{\infty}f(x)dx=\int_{0}^{a}f(x)dx+\int_{a}^{\infty}f(x)dx<\int_{0}^{a}f^{*}(x)dx+\int_{a}^{\infty}f^{*}(x)dx=1.$
Contradiction arises. Suppose $a>t_{0}.$ Note that $\int_{a}^{\infty}f(x)dx\leq ca^{-\lambda}=\int_{a}^{\infty}f^{*}(x)dx$.
Thus, $\int_{0}^{\infty}f(x)dx=\int_{0}^{a}f(x)dx+\int_{a}^{\infty}f(x)dx<\int_{0}^{a}f^{*}(x)dx+\int_{a}^{\infty}f^{*}(x)dx=1.$
Contradiction arises. Therefore, $f\succ f^{*}$.
\end{IEEEproof}
We now present the proof of Theorem \ref{thm:p_inf}.
\begin{IEEEproof}
[Proof of Theorem \ref{thm:p_inf}]Let $f$ be the pdf of the distribution
$P$, and $X$ be a random variable following the distribution of
$P$. Consider the distribution of $\lceil X\rceil$. We have, for
integer $x\ge0$,
\begin{align*}
\mathbf{P}(\lceil X\rceil>x) & =\mathbf{P}\left(X>x\right)\\
 & \le cx^{-\lambda}.
\end{align*}
By Theorem \ref{thm:p_int}, we have 
\[
\mathbf{E}(L(W_{\mathrm{int}}))\le\dfrac{50c\lambda n^{\frac{1}{\lambda}}\log(\sqrt{n}+1)}{\lambda-1}.
\]
Consider $W_{i}$ for $i\ge1$. By Theorem \ref{thm:p_01},
\begin{align*}
 & \mathbf{E}(L(W_{i}))\\
 & \leq\mathbf{E}\left(\mathbf{1}\{n_{i}\ge1\}\cdot92\sqrt{n_{i}f_{i}(0)}(\log(\sqrt{n_{i}f_{i}(0)}+1))\right)\\
 & =\mathbf{E}\left(92\sqrt{n_{i}f_{i}(0)}(\log(\sqrt{n_{i}f_{i}(0)}+1))\right)\\
 & \overset{(a)}{\leq}92\sqrt{\mathbf{E}(n_{i})f_{i}(0)}(\log(\sqrt{\mathbf{E}(n_{i})f_{i}(0)}+1))\\
 & \overset{(b)}{=}92\sqrt{nf(i-1)}(\log(\sqrt{nf(i-1)}+1)),
\end{align*}
where $(a)$ is by Jensen's inequality and the concavity of $\sqrt{t}\log(\sqrt{t}+1)$,
and $(b)$ is by 
\begin{align*}
\mathbf{E}(n_{i})f_{i}(0) & =\left(n\int_{i-1}^{i}f(t)dt\right)\frac{f(i-1)}{\int_{i-1}^{i}f(t)dt}\\
 & =nf(i-1).
\end{align*}
Hence, we have
\begin{align*}
 & \sum_{i=1}^{\infty}\mathbf{E}(L(W_{i}))\\
 & \le\sum_{i=1}^{\infty}92\sqrt{nf(i-1)}(\log(\sqrt{nf(i-1)}+1))\\
 & \leq92\sqrt{nf(0)}(\log(\sqrt{nf(0)}+1)\\
 & \ \ \ \ +\int_{0}^{\infty}92\sqrt{nf(x)}\left(\log(\sqrt{nf(x)}+1)\right)dx.
\end{align*}
Consider the term $\int_{0}^{\infty}72\sqrt{nf(x)}\left(\log(\sqrt{nf(x)}+1)\right)dx$.
By Lemma \ref{lem:maj_concave}, Lemma \ref{lem:power_bd} and the
concavity of $\sqrt{t}\log(\sqrt{t}+1)$, 

\begin{align*}
 & \int_{0}^{\infty}92\sqrt{nf(x)}\left(\log(\sqrt{nf(x)}+1)\right)dx\\
 & \le\int_{0}^{\infty}92\sqrt{nf^{*}(x)}\left(\log(\sqrt{nf^{*}(x)}+1)\right)dx\\
 & =\int_{0}^{t_{0}}92\sqrt{nf^{*}(x)}\left(\log(\sqrt{nf^{*}(x)}+1)\right)dx\\
 & \ \ \ \ +\int_{t_{0}}^{\infty}92\sqrt{nf^{*}(x)}\left(\log(\sqrt{nf^{*}(x)}+1)\right)dx\\
 & \leq92t_{0}\sqrt{nf^{*}(0)}\left(\log(\sqrt{nf^{*}(0)}+1\right)\\
 & \ \ \ \ +92\int_{t_{0}}^{\infty}\sqrt{nc\lambda x^{-\lambda-1}}\left(\log(\sqrt{nf^{*}(0)}+1)\right)dx\\
 & \overset{(c)}{\leq}92t_{0}\sqrt{n}\left(\log(\sqrt{n}+1\right)\\
 & \ \ \ \ +92\int_{t_{0}}^{\infty}\sqrt{nc\lambda x^{-\lambda-1}}\left(\log(\sqrt{n}+1)\right)dx\\
 & =92\left(c(\lambda+1)\right)^{\frac{1}{\lambda}}\sqrt{n}\log(\sqrt{n}+1)\\
 & \ \ \ \ +\frac{184\sqrt{nc\lambda}\log(\sqrt{n}+1)}{(\lambda-1)\left(c(\lambda+1)\right)^{\frac{\lambda-1}{2\lambda}}}\\
 & \leq92\left(c(\lambda+1)\right)^{\frac{1}{\lambda}}\sqrt{n}\log(\sqrt{n}+1)\\
 & \ \ \ \ +\frac{184(c(\lambda+1))^{\frac{1}{2\lambda}}\sqrt{n}\log(\sqrt{n}+1)}{(\lambda-1)}\\
 & \leq\frac{276\left(c(\lambda+1)\right)^{1/\lambda}\sqrt{n}\log(\sqrt{n}+1)}{\min(\lambda-1,1)},
\end{align*}
where $(c)$ follows from $f^{*}(0)=c\lambda\left(c(\lambda+1)\right)^{-1-\frac{1}{\lambda}}\leq1$.
Hence,

\begin{align*}
 & \sum_{i=1}^{\infty}\mathbf{E}(L(W_{i}))\\
 & \leq92\sqrt{nf(0)}(\log(\sqrt{nf(0)}+1)\\
 & \ \ \ \ +\frac{276\left(c(\lambda+1)\right)^{1/\lambda}\sqrt{n}\log(\sqrt{n}+1)}{\min(\lambda-1,1)}.
\end{align*}

Therefore, the expected codeword length

\begin{align*}
 & \mathbf{E}(L(W_{\mathrm{int}}))+\sum_{i=1}^{\infty}\mathbf{E}(L(W_{i}))\\
 & \leq\dfrac{50c\lambda n^{\frac{1}{\lambda}}\log(\sqrt{n}+1)}{\lambda-1}+92\sqrt{nf(0)}(\log(\sqrt{nf(0)}+1)\\
 & \ \ \ \ +\frac{276\left(c(\lambda+1)\right)^{1/\lambda}\sqrt{n}\log(\sqrt{n}+1)}{\min(\lambda-1,1)}\\
 & \leq\dfrac{418c(\lambda+1)\max(n^{1/\lambda},\sqrt{n})\max(\sqrt{f(0)},1)}{\min(\lambda-1,1)}\\
 & \ \ \ \ \times\log(\sqrt{n\max(f(0),1)}+1).
\end{align*}

Therefore, the expected codeword length grows sublinearly.
\end{IEEEproof}
Our coding scheme described in Definition \ref{def:The-coding-scheme, =00005B0,inf)}
uses the coding scheme in Defintion \ref{def:scheme_p_01} as a building
block. Therefore, if we use Theorem \ref{thm:p_01} in our analysis,
then even if we assume a stronger tail bound, such as exponential
tail bound $\text{\ensuremath{\mathbf{P}}}(X>x)\leq ce^{-\lambda x},$
the order of the growth of expected codeword length cannot be better
than $O(\sqrt{nf(0)}(\log(\sqrt{nf(0)}+1)))$.

\medskip{}

\section{Conclusion and Discussion}

In this paper, we introduced a new problem in channel simulation called
the multiple-output channel simulation. We also describe encoding
schemes for three classes of probability distributions and show that
the growth rate of the expected codeword length is sub-linear in $n$
when a power bound or exponential bound is satisfied. An application
of multiple-outputs channel simulation is the compression of probability
distributions. 

We list some potential extenstions of our result. First, it may be
possible to generalize the result to more classes of probability distributions,
such as unimodal distributions over the real line. Second, since this
paper only focus on upper bounds of our codeword length, lower bounds
may also be derived in the future in order to show tightness. Third,
we may also consider the case where common randomness is available
to the encoder and decoder.

\section{Acknowledgment}

The authors acknowledge support from the Direct Grant for Research,
The Chinese University of Hong Kong. 

\bibliographystyle{ieeetr}
\bibliography{ref}

\end{document}